\documentclass[useAMS,usenatbib]{mn2e}
 
\usepackage{graphicx}

\title[Spectroscopy of Hyades L dwarf candidates]
{Spectroscopy of Hyades L dwarf candidates
\thanks{Based on observations collected with telescopes
operated on the island of La Palma in the Spanish Observatorio del
Roque de los Muchachos of the Instituto de Astrof\'isica de Canarias.}}
\author[N. Lodieu]{N. Lodieu$^{1,2}$\thanks{E-mail: nlodieu@iac.es},
S. Boudreault$^{1,2,3}$, and V. J. S. B\'ejar$^{1,2}$ \\
$^{1}$ Instituto de Astrof\'isica de Canarias (IAC), C/ V\'ia L\'actea s/n, 
E-38200 La Laguna, Tenerife, Spain \\
$^{2}$ Departamento de Astrof\'isica, Universidad de La Laguna (ULL),
E-38206 La Laguna, Tenerife, Spain \\
$^{3}$ GEPI, Observatoire de Paris, CNRS, Universit\'e Paris Diderot; 
5 Place Jules Janssen, F-92190 Meudon, France \\
}
\begin{document}

\date{Accepted \today. Received \today; in original form \today}

\pagerange{\pageref{firstpage}--\pageref{lastpage}} \pubyear{2005}

\maketitle

\label{firstpage}

%
%
\begin{abstract}
We present the results of photometric, astrometric, and spectroscopic follow-up of L dwarf
candidates identified in the Hyades cluster by \citet{hogan08}. We obtained low-resolution
optical spectroscopy with the OSIRIS spectrograph on the Gran Telescopio de Canarias for 
all 12 L dwarf candidates as well as new $J$-band imaging for a subsample of eight to confirm 
their proper motion. We also present mid-infrared photometry from the Wise Field Infrared
Survey Explorer (WISE) for the Hyades L and T dwarf candidates and estimate their spectroscopic
distances, effective temperatures, and masses.
We confirm the cool nature of several L dwarf candidates and confirm astrometrically
their membership, bridging the gap between the
coolest M dwarfs and the two T dwarfs previously reported in the Hyades cluster. 
These members represent valuable spectral templates at an age of 625 Myr
and slightly super solar metallicity (Fe/H\,=\,$+$0.13).
We update the Hyades mass function across the hydrogen-burning limit and in the
substellar regime. We confirm a small number numbers of very-low-mass members
below $\sim$0.1 M$_{\odot}$ belonging to the Hyades cluster.
\end{abstract}


%
%
\section{Introduction}
\label{Hyades_dL:intro}

The shape of the Initial Mass Function (IMF) is of prime importance to understand 
the processes responsible for the formation of stars and brown dwarfs. The form of
the IMF across the stellar/substellar limit is of particular interest to assess
whether stars and brown dwarfs represent two distinctive populations
\citep{thies07,thies08} or have properties differing due to dynamical
interactions \citep{parker14}. The definition and first estimate of the IMF 
was presented by \citet{salpeter55} and updated by \citet{miller79} and 
\citet{scalo86}. Our current knowledge of the IMF suggests that it is best 
fit by a lognormal form \citep{chabrier03,chabrier05a} or a combination of 
power laws \citep{kroupa02,kroupa13}, with little variations across a large
range of masses and environments \citep[see review by][]{bastian10}.
Studying the form and shape of the bottom-end of the mass function in relatively 
old clusters like the Hyades, where low-mass members may have been evaporated by 
dynamical evolution, is of interest to study the intrinsic evolution of individual 
brown dwarfs and how the stellar and substellar populations evolve.
Looking at the shape of the IMF at different metallicities is also important
to probe its dependance with environment.

The Hyades cluster has a mean distance of 46.3$\pm$0.3 pc, a significant
proper motion ranging between 74 and 140 mas/yr, a tidal radius of $\sim$10 pc, 
and a core radius of 2.5--3.0 pc, according to the analysis of the Hipparcos catalogue
by \citet{perryman98}. The age of the cluster is estimated to be 625$\pm$50 Myr
based on reproducing the observed cluster sequence with model isochrones which 
include convective overshooting \citep{maeder81}, although a larger range
in age cannot be discarded \citep{eggen98a}.
The metallicity of the Hyades high-mass stars appears slightly super-solar,
with values between 0.127$\pm$0.022 and 0.14$\pm$0.1 \citep{boesgaard90,cayreldestrobel97,grenon00}
although a more recent work by  \citet{gebran10} suggests a mean metallicity
close to solar ([Fe/H]\,=\,0.05$\pm$0.05).
Despite being one of the closest and best studied open cluster,
the Hyades still hold secrets regarding its membership, dynamics, and evolution. 
The majority of surveys have focused on small patches of the sky near the cluster 
centre to identify new members \citep{hanson75,leggett89,reid93,stauffer94a,stauffer95,reid97b,reid99a,dobbie02c}. 
Some larger area surveys have been undertaken in the past \citep[e.g.][]{reid92}, 
recently updated by the study of \citet{roeser11} using the PPMXL catalogue \citep{roeser10}.
However, this search was limited in magnitude to $V$$\sim$20 mag, yielding a
catalogue of 364 stars within the tidal radius of the cluster, down to 
0.2 M$_{\odot}$ corresponding to a spectral type of M4\@. 
This work was recently extended by \citet{goldman13} down to lower masses
($\sim$0.09 M$_{\odot}$) by combining PPMXL with PanStarrs \citep{kaiser02}
to derive the cluster mass function in the low-mass regime.
\citet{gizis99b} attempted to select photometrically L dwarf candidates 
in the Hyades from a photometric search in the Two Micron All Sky Survey
\citep[2MASS;][]{cutri03,skrutskie06} but did not confirm any of the candidate as
spectroscopic member. \citet{hogan08} conducted a similar photometric and
astrometric work cross-correlating 2MASS and the UKIRT Infrared Deep Sky Survey 
\citep[UKIDSS;][]{lawrence07} Galactic Clusters Survey (GCS), yielding a sample 
of 12 L dwarf candidates, four of them recently confirmed as ultracool dwarfs 
by near-IR spectroscopy \citep{casewell14a}.
At the low-mass end of the mass function, \citet{bouvier08a} identified
spectroscopically the first two T dwarfs in the Hyades, with masses estimated
to $\sim$0.05 M$_{\odot}$ according to theoretical isochrones \citep{baraffe98,chabrier00c}.

In this paper, we present an astrometric, photometric, and spectroscopic follow-up
of L dwarf candidates in the Hyades open cluster published by \citet{hogan08}
to assess their membership.
In Section \ref{Hyades_dL:phot_astrom} we describe the photometric observations
and its associated data reduction.
In Section \ref{Hyades_dL:spectro} we present the optical spectroscopy 
carried out with the Gran Telescopio de Canarias (GTC).
In Section \ref{Hyades_dL:memb} we derive the spectral types of the Hyades
candidates presented in \citet{hogan08} and discuss their membership
based on our photometric, astrometric, and spectroscopic observations.
In Section \ref{Hyades_dL:physical_properties} we derive physical properties
of the confirmed L dwarfs, including spectroscopic distances, effective
temperatures, and masses based on state-of-the-art models.
In Section \ref{Hyades_dL:MF} we discuss the shape of the luminosity and mass
functions and compare them to the Praesepe cluster whose age is comparable to
the Hyades.

%
%
\section{Photometric and astrometric follow-up}
\label{Hyades_dL:phot_astrom}
\subsection{Near-infrared imaging}
\label{Hyades_dL:phot_obs}

We conducted $J$-band imaging follow-up with the Long-slit Intermediate Resolution Infrared Spectrograph
\citep[LIRIS;][]{manchado98} at the Cassegrain of the William Herschel Telescope (WHT) located on the
Observatorio del Roque de Los Muchachos en La Palma (Canary Islands). We carried out this project
on 13 January 2014 as a back-up program during an unrelated run due to the poor transparency (thin
clouds passing by) and variable seeing poorer than 1.3 arcsec because of the
brightness of the Hyades targets. We observed eight of the 12 L dwarf candidates of
\citet{hogan08}: Hya03--08, Hya10, and Hya12\@.

LIRIS is equipped with a 1024$\times$1024 pixel HAWAII detector sensitive to near-infrared wavelengths
(0.8--2.5 microns) with a pixel scale of 0.25 arcsec, giving a field-of-view of 4.7 by 4.7 arcmin across.
The instrument is equipped with a large set of filters, including the $J$ broad-band filter
used for our photometric follow-up. We employed a 9-point dithering pattern with on-source individual
integrations of 5 sec repeated one or twice with a small jitter offset, yielding a total on-source integration
of 5 min or 10 min. We obtained dome flats during the afternoon preceding the observations.
We did not observe any photometric standard stars because our fields are covered by
2MASS, providing us with magnitudes calibrated to the 2MASS system. We will use 
these observations to complement the $K$-band photometry of UKIDSS and provide an additional 
epoch to the 2MASS and UKIDSS epochs.

%
%
\begin{table*}
 \centering
 \caption[]{Log of the WHT LIRIS and GTC OSIRIS photometric and spectroscopic
observations for the Hyades L dwarf candidates published by \citet{hogan08}.
Coordinates are from our LIRIS observations taken on 13 January 2014 (in J2000),
except for the four candidates not observed photometrically which are taken 
from \citet{hogan08}. ``Nb'' stands for the numbers of exposures at a given
position.
}
{\scriptsize
 \begin{tabular}{@{\hspace{0mm}}c @{\hspace{2mm}}c @{\hspace{2mm}}c | @{\hspace{2mm}}c @{\hspace{2mm}}c @{\hspace{2mm}}c @{\hspace{2mm}}c @{\hspace{2mm}}c @{\hspace{2mm}}c | @{\hspace{2mm}}c @{\hspace{2mm}}c @{\hspace{2mm}}c @{\hspace{2mm}}c @{\hspace{2mm}}c | @{\hspace{2mm}}c@{\hspace{0mm}}}
 \hline
 \hline
ID  & R.A.    &     Dec       &   \multicolumn{6}{|c}{Photometry} & \multicolumn{5}{|c}{Spectroscopy}  \cr
 \hline
    &          &                     &   Tel/Instr.\ & Date   & ExpT & Nb & Seeing & $J$ (errJ) & Tel/Instr.\ & Date & ExpT & Nb & Grism &  \cr
\hline
    & hh:mm:ss.ss & ${^\circ}$:$'$:$''$ & & ddmmyyyy & sec &  &  arcsec & mag &  & ddmmyyyy & sec &  &  &  \cr
 \hline
Hya01 & 04:20:24.50 & $+$23:56:13.0 &    ---    &   ---    &    &    &     &  ---   & GTC OSIRIS & 23102013 &  600 & 1 & R500R & M8.5  \cr
Hya02 & 03:52:46.30 & $+$21:12:33.0 &    ---    &   ---    &    &    &     &  ---   & GTC OSIRIS & 25102013 &  600 & 1 & R300R & L1.5  \cr
Hya03 & 04:10:24.01 & $+$14:59:10.3 & WHT LIRIS & 13012014 & 10 & 18 & 1.3 & 15.647 (0.076) & GTC OSIRIS & 16102013 &  300 & 1 & R500R & L0.5 \cr
Hya04 & 04:42:18.59 & $+$17:54:37.3 & WHT LIRIS & 13012014 & 10 & 18 & 1.5 & 15.595 (0.044) & GTC OSIRIS & 16102013 &  300 & 1 & R500R & M9.5  \cr
Hya05 & 03:58:43.06 & $+$10:39:39.6 & WHT LIRIS & 13012014 & 10 & 18 & 1.4 & 15.729 (0.071) & GTC OSIRIS & 16102013 &  300 & 1 & R500R & M5    \cr
Hya06 & 04:22:05.22 & $+$13:58:47.3 & WHT LIRIS & 13012014 & 10 & 18 & 1.7 & 15.462 (0.081) & GTC OSIRIS & 16102013 &  300 & 1 & R500R & M9.5  \cr
Hya07 & 04:39:29.17 & $+$19:57:34.6 & WHT LIRIS & 13012014 & 10 & 18 & 1.8 & 15.956 (0.073) & GTC OSIRIS & 16102013 &  300 & 1 & R500R & M3    \cr
Hya08 & 04:58:45.75 & $+$12:12:34.1 & WHT LIRIS & 13012014 & 10 & 18 & 1.3 & 15.463 (0.044) & GTC OSIRIS & 16102013 &  300 & 1 & R500R & L0.5  \cr
Hya09 & 04:46:35.40 & $+$14:51:26.0 &    ---    &   ---    &    &    &     &  ---   & GTC OSIRIS & 22032014 & 1800 & 1 & R500R & L2.0   \cr
Hya10 & 04:17:33.97 & $+$14:30:15.4 & WHT LIRIS & 13012014 & 10 & 18 & 1.4 & 16.506 (0.059) & GTC OSIRIS & 26102013 &  600 & 4 & R300R & L1.0  \cr
Hya11 & 03:55:42.00 & $+$22:57:01.0 &     ---   &    ---   &    &    &     &  ---   & GTC OSIRIS & 23102013 &  600 & 1 & R500R & L1.5  \cr
Hya12 & 04:35:43.02 & $+$13:23:44.8 & WHT LIRIS & 13012014 & 10 & 18 & 1.3 & 16.778 (0.069) & GTC OSIRIS & 26102013 &  600 & 4 & R300R & L3.5 \cr
 \hline
 \label{tab_Hyades_dL:log_obs}
 \end{tabular}
}
\end{table*}

\subsection{Data reduction}
\label{Hyades_dL:phot_DR}

We carried out the data reduction of the LIRIS images on the fly on the mountain with the LIRIS
pipeline\footnote{www.ing.iac.es/Astronomy/instruments/liris/liris\_ql.html}. This procedure is
fairly automatised and consists of subtracting the sky made from the dithered images to each
individual image of the target. The pipeline also includes a correction for flat field (median
image made of the median dome flats taken during the afternoon), vertical gradient observed on the 
detector, and the geometrical distortion. The corrected images obtained per object are then 
aligned and combined to provide a final image taking into account the random offsets between 
the dithered and jittered positions.

The pipeline-reduced combined images include a rough astrometric calibration, but there is
a significant offset ($\sim$20 arcsec) with the world coordinate system. To obtain an accurate 
astrometric calibration of the LIRIS images, we have made use of the Graphical Astronomy and 
Image Analysis Tool (GAIA)\footnote{http://star-www.dur.ac.uk/~pdraper/gaia/gaia.html}. 
We employed the astrometric facility under the image analysis tool by tweaking the
existing calibrations of the UKIDSS GCS \citep{lawrence07}. We shifted all UKIDSS GCS 
detections brighter than $K$\,=\,17.8 mag to their positions on the LIRIS images and 
removed objects lying on the edge of the detector to optimize the six-parameter polynomial
fit (order of two) of the astrometric solution. We obtained rms errors on the astrometric 
fit of 0.63--0.93 pixel, i.e.\ 0.16--0.23 arcsec, based on samples of a few tens of stars.

We used the aforementioned astrometrically-calibrated images to derive the photometry.
We derived instrumental magnitudes using the photometric analysis tools in GAIA, which makes
use of SEXtractor \citep{bertin96}. We employed a fixed aperture radius of 8 pixels (=2 arcsec)
for all images, slightly larger than the full-width-half-maximum which oscillated between 
1.2 and 1.9 arcsec during the observations of the Hyades cluster members. 
We inferred photometric zero points for each LIRIS image from point sources
brighter than $J_{\rm 2MASS}$\,=\,15.9 mag, equivalent to a 10$\sigma$ detection in 2MASS\@.
We point out that this process cannot be carried out with the UKIDSS GCS catalogue because
only $K$-band photometry is available towards the Hyades cluster.
This limit in magnitude yielded photometric zero points in the 22.398--23.024 magnitude range 
with uncertainties less than six percent, based on five to 23 stars in the LIRIS field-of-view. 
We list the final $J$-band magnitudes of Hya03--08, Hya10, and Hya12 and the
details of the observations in Table \ref{tab_Hyades_dL:log_obs}. We find good agreement 
with the 2MASS magnitudes quoted by \citet{hogan08} within the photometric uncertainties.

%
%
%
\section{Optical spectroscopy}
\label{Hyades_dL:spectro}
\subsection{Spectroscopic observations}
\label{Hyades_dL:spectro_Obs}

We obtained low-resolution (R\,$\sim$\,200 or 400) optical spectroscopy with the OSIRIS
\citep[Optical System for Imaging and low-intermediate Resolution Integrated Spectroscopy;][]{cepa00}
mounted on the 10.4-m GTC telescope in La Palma over several nights. OSIRIS is equipped with
two 2048$\times$4096 Marconi CCD42-82 detectors offering a field-of-view
approximately 7$\times$7 arcmin with an unbinned pixel scale of 0.125 arcsec.
We observed the Hyades L dwarf candidates as part of a run in visitor mode
(GTC27-13B) and two filler programmes in service mode (GTC65-13B; GTC5-14A; PI Lodieu). 
We observed three L dwarf candidates Hya02, Hya10, and Hya12 during our visitor 
mode run as back-up targets due to the poor transparency and bad seeing conditions. 
We took spectra for the remaining nine L dwarf candidates under variable conditions 
as part of our filler programmes, which accepts seeing worse than 1.5 arcsec, bright time, 
and cirrus. The slit was set to 1 arcsec or 1.5 arcsec depending on the seeing.
The log of the observations are in Table \ref{tab_Hyades_dL:log_obs}.
We observed the spectro-photometric standards, G191-B2B \citep{vanLeeuwen07,gianninas11}
and Hilt\,600 \citep{hog00,pancino12} with both gratings. We obtained an additional
spectrum of the standards with the $Z$ filter to correct for the second-order contamination 
beyond 9000\AA{} which affects those low-resolution gratings \citep[see also][]{zapatero14a}.
Bias frames, dome flat fields, and Neon, Xenon, and HgAr arc lamps were observed
by the observatory staff during the afternoon preceding the observations.

We have also made use of optical spectral templates of M8--L4 dwarfs taken with 
GTC OSIRIS taken as part of programmes GTC66-12B and GTC62-13B (PI Boudreault).
We used the grism R300R and a slit of one arcsec with a single on-source integration
between 300\,sec and 900\,sec scaled according to the magnitude.
We emphasise that the settings used for these spectral templates closely resembles
the ones used for the Hyades targets, yielding similar spectral resolution.
We considered the following field M and L dwarfs classified in the optical:
LP\,213-68 \citep[M8;][]{reid02c}, 2MASS\,J02512220$+$2521236 \citep[M9;][]{hawley02}
2MASP\,J0345432$+$254023 \citep[L0;][]{kirkpatrick97,kirkpatrick99,dahn02,knapp04},
SDSS\,J163050.01$+$005101.3 \citep[L1;][]{hawley02,jameson08a}, 
2MASSW\,J0030438$+$313932 \citep[L2;][]{kirkpatrick99,jameson08a}, 
2MASSW\,J0355419+225702 \citep[L3;][]{kirkpatrick99}, and
2MASSW\,J1155009$+$230706 \citep[L4;][]{kirkpatrick99,jameson08a}. 

\subsection{Data reduction}
\label{Hyades_dL:spectro_DR}

\begin{figure*}
  \centering
  \includegraphics[width=\linewidth, angle=0]{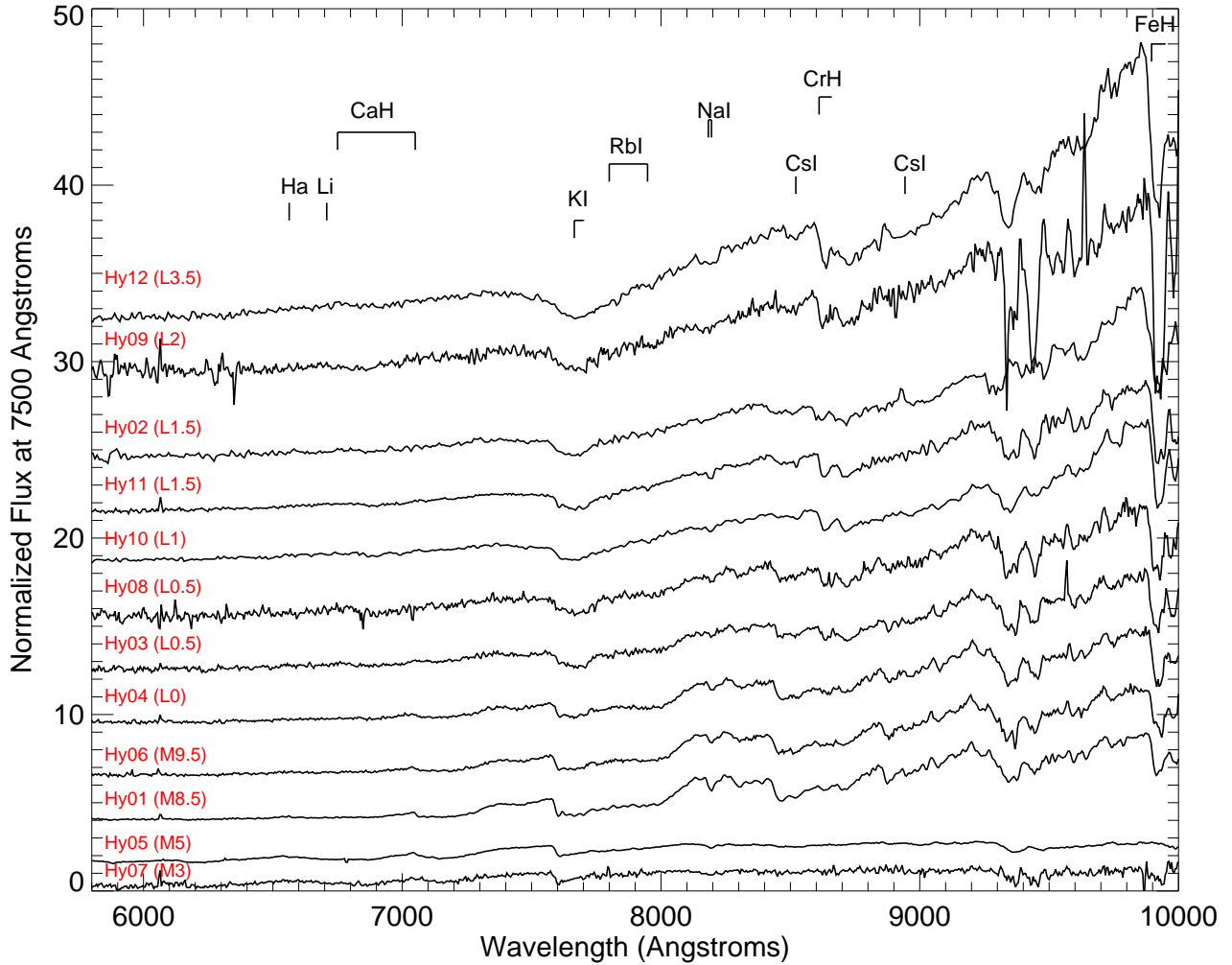}
  \caption{Low-resolution optical spectra for 12 Hyades L dwarf candidates
  from \citet{hogan08} obtained with the R300R and R500R gratings on the 
  GTC/OSIRIS spectrograph. Some important absorption bands and atomic lines
  affecting the spectral energy distribution of M and L dwarfs are labelled.}
  \label{fig_Hyades_dL:OSIRIS_opt_spec}
\end{figure*}

We reduced the OSIRIS optical spectrum following a standard methodology under the IRAF environment
\citep{tody86,tody93}. We substracted the raw spectrum by a median-combined bias and
divided by a normalised median-combined dome flat field taken during the
afternoon. We extracted optimally a 1D spectrum from the 2D image and
calibrated that spectrum in wavelength with the lines from the combined arc lamp.
We calibrated the spectra of the Hyades in flux with the spectro-photometric standard
corrected for the second-order contamination. The spectral templates,
which were taken prior to summer 2013, were observed solely with the grating and
without the Z filter, hence, were not corrected for the second-order contamination. 
As a consequence, the flux calibration of these spectral standards
is only trustworthy up to 900 nm. We combined the template
spectra to create references for each half subtype (L0.5, L1.5, L2.5, and L3.5).
The GTC spectra for 12 Hyades L dwarf candidates, normalised at 7500\,\AA{},
are displayed in Figure \ref{fig_Hyades_dL:OSIRIS_opt_spec}.

%
%
%

%
%
\begin{figure}
  \centering
  \includegraphics[width=\linewidth, angle=0]{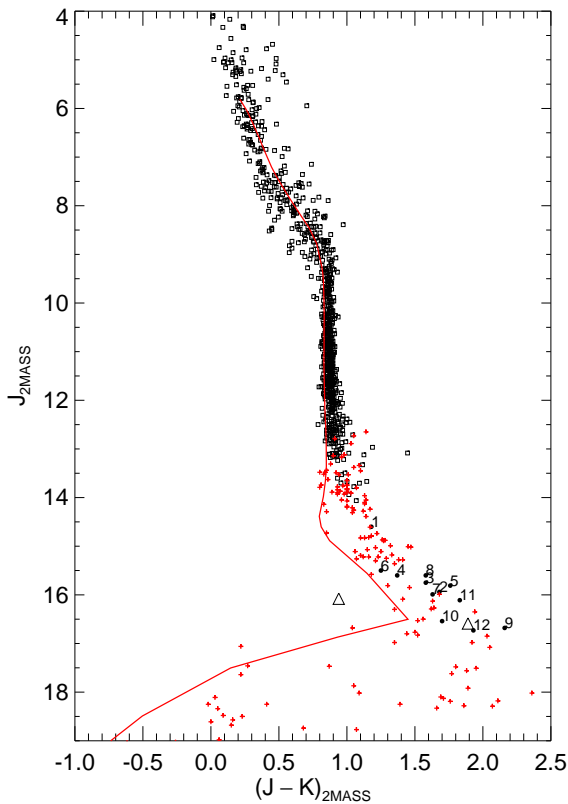}
  \caption{($J-K_{s}$,$J$) colour-magnitude diagram for known Hyades stellar members 
from \citet{goldman13} marked as open squares. Overplotted are the 12 L dwarf candidates
(filled circles with ID numbers) from \citet{hogan08}. Candidates numbers 5 and 7
are rejected as members by our spectroscopic follow-up.
For completeness, we added the two T dwarfs confirmed spectroscopically by 
\citet{bouvier08a} as open triangles.
Overplotted as red crosses are ultracool dwarfs with parallaxes from \citet{dupuy12}
as well as the 600 Myr BT-Settl isochrone (red line) shifted at the distance 
of the Hyades.
}
  \label{fig_Hyades_dL:CMD_JKJ}
\end{figure}
%

%
%
\begin{figure}
  \centering
  \includegraphics[width=\linewidth, angle=0]{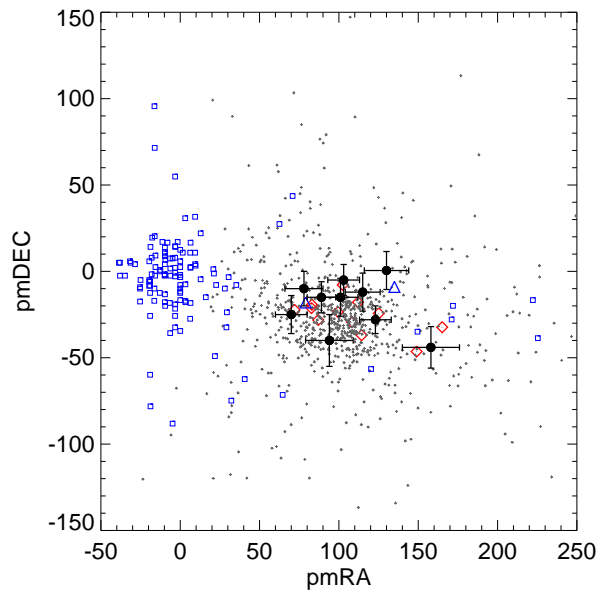}
  \caption{Vector point diagram for Hyades L dwarf candidates.
Large filled dots are proper motions with their associated error bars
derived from the LIRIS vs UKIDSS GCS DR10 cross-match presented in this paper, 
while large red open diamonds are measurements from \citet{hogan08}. 
Small open blue square represent the dispersion of all point sources in the LIRIS 
fields. Overplotted as grey symbols are all Hyades members from \citet{goldman13}. 
T dwarfs from \citet{bouvier08a} are displayed as large blue open triangles.
}
  \label{fig_Hyades_dL:plot_VPD}
\end{figure}
%

%
%
\section{Membership of L dwarf candidates}
\label{Hyades_dL:memb}
\subsection{Photometric membership}
\label{Hyades_dL:memb_phot_NIR}

\citet{hogan08} selected 12 L dwarf candidates photometrically and astrometrically.
We obtained $J$-band photometry for eight of these 12 L dwarfs (Hya03--08, Hya10, and Hya12)
and confirmed their magnitudes compared to 2MASS within the photometric errors. 
In Fig.\ \ref{fig_Hyades_dL:CMD_JKJ} we plot the
($J-K_{s}$,$J$) colour-magnitude depicting known Hyades members \citet{goldman13} 
along with the 12 L dwarf candidates (dots with their associated identification number) 
and the two T dwarfs \citep[open triangles;][]{bouvier08a}. We overplotted the
600 Myr isochrones from the BT-Settl models  \citep{allard12} as well as ultracool
dwarfs with parallaxes from \citet*{dupuy12}. We note that we removed subdwarfs from
their list. We can see that the L dwarf candidates bridge the gap between 
Hyades M and T dwarfs, adding credit to their membership.

\subsection{Astrometric membership}
\label{Hyades_dL:memb_astrom}

To carry out the proper motion analysis, we made use of UKIDSS images as first epoch 
(images typically taken between 9 October and 27 December 2005) and LIRIS images as 
a second epoch. These data provide a baseline of about eight years. 
For Hya02 and Hya11 we made use of two epochs of UKIDSS separated by about five years
for the proper motion determination because we did not obtain LIRIS imaging.
We estimated the proper motion by 
comparison of the relative positions of the targets with respect to about 15--30 reference 
stars located in a field of 4 arcmin by 4 arcmin around the candidate position. 
We used the {\tt{GEOMAP}} routine within the IRAF enviroment to convert pixel coordinates 
from the first to the second epoch. We employed a general transformation with a polynomial 
function of order three. We converted the resulting pixel displacements of our targets 
into proper motions using the astrometric solution given by the UKIDSS images. 
We estimate the uncertainties in the proper motion by summing quadratically the errors 
in the centroids, which are of the order of 1/3 and 1/30 of pixel in the LIRIS and UKIDSS 
images, respectively, and the standard deviation of the transformation of reference
stars. We indicate the resulting proper motions and their uncertainties in
Table \ref{tab_Hyades_dL:table_results}. The mean proper motion of the L dwarf 
candidates is ($\mu{_\alpha}\cos{\delta}$,$\mu{_\delta}$)\,=\,(106.10,$-$19.35) mas/yr
with an intrinsic dispersion (after removing proper motion errors) of 
25.5 mas/yr and 7.4 mas/yr in right ascension and declination,
respectively (total is 26.5 mas/yr). For comparison, the total intrinsic dispersion of
Hyades members from \citet{goldman13} in the area covered by \citet{hogan08}
with masses in the 2.6--2.0 M$_{\odot}$,
2.0--1.0 M$_{\odot}$, 1.0--0.5 M$_{\odot}$, and 0.5--0.2 M$_{\odot}$ are
17.96 mas/yr, 19.67 mas/yr, 23.82 mas/yr, and 26.89 mas/yr, respectively.

In Figure \ref{fig_Hyades_dL:plot_VPD} we plot the measured proper motions of the 
eight L dwarf candidates (black dots) observed with WHT/LIRIS in a 
vector point diagram to further assess their membership. We overplotted in
that figure the known Hyades members published by \citet{goldman13} with small
grey crosses. We can observe the large dispersion of the members, centered
around (100,$-$25) mas/yr. We also added the proper motions measured for point 
sources lying in the LIRIS fields-of-view blue open squares. On the one 
hand, we confirm that most objects lie around (0,0) in the vector point 
diagram, except for several cases which we cannot discard as members of the 
Hyades. On the other hand, we estimate a mean astrometric error representing
the quadratic sum of the of the centroid errors and standard deviation of the 
the order of 15 mas/yr, based on the dispersion of the open squares.

Moreover, we confirm the proper motions published in Table 2 of \citet{hogan08}.
Overall, we conclude that the eight L dwarf candidates followed-up
photometrically with LIRIS have a proper motion consistent with more massive 
members of the Hyades cluster.

\subsection{Spectroscopic membership}
\label{Hyades_dL:memb_spectro}

We have investigated further the nature of the 12 L dwarf candidates with our
GTC/OSIRIS spectroscopic follow-up. To assign the spectral types to our 
candidates, we opted for the direct comparison with spectral templates. 
For spectral types earlier than (or equal to) L0 (Hya01, Hya04, Hya05, 
and Hya07), we made use of the Sloan 
Digital Sky Survey \citep[SDSS;][]{york00} spectroscopic database \citet{bochanski07a}.
This database contains a repository of good-quality M0--L0 spectra spanning 
the 380--940 nm wavelength range at a resolution
of 2000\@. All spectra are wavelength- and flux-calibrated and corrected for
telluric absorption. M dwarfs are classified based on the Hammer classification
scheme \citep{covey07} which uses the spectral energy distribution of stars over
the 0.3--2.5 micron range using photometry from SDSS and 2MASS\@.
We added sub-types to this list of templates to provide a complete list of
spectral templates with optical spectral types accurate to one sub-type.

For spectral types later than L0 (Hya02--03, Hya08--12), we employed 
spectral templates observed
with GTC OSIRIS at a similar spectral resolution for direct comparison
(Section \ref{Hyades_dL:spectro_Obs}). We have spectral templates for each
half subtype, from L0 up to L4\@. We note that the GTC optical spectra of
our M8, M9, and L0 spectral templates agree closely with the Sloan templates.
Our uncertainty on the spectral types is 0.5 subtype.

\citet{casewell14a} derived spectral types for five candidates in common to
our work, based on low-resolution near-infrared spectra.
They rejected Hya02 as a Hyades members whereas we can clearly see
that its spectrum looks like a L dwarf with a spectral type of L0.5
(Fig.\ \ref{fig_Hyades_dL:OSIRIS_opt_spec}). We classify two objects (Hya01
and Hya04) roughly two spectral types earlier than \citet{casewell14a},
M8.5 vs L0.5 and L0 vs L2--L3, respectively. Our uncertainty on the
spectral type is lower than the ones from the near-infrared
classification of \citet{casewell14a}. This trend of earlier spectral 
types derived from optical spectra vs near-infrared spectra has already been 
seen in young dwarfs \citep[e.g.][]{lodieu05b}. Our spectral types for Hya03
and Hya06 agree with those of \citet{casewell14a}, L0.5 vs M8--L0.5 and
M9.5 vs M8--L2, but our uncertainty on the spectral type is accurate
to half a subclass, better than the near-infrared spectral types of
\citet{casewell14a}.

To summarise, we found two M dwarf contaminants (Hya05 and Hya07), 
three late-M dwarfs (Hya01, Hya04, and Hya06), and seven L dwarfs 
(spectral types between L0 and L3.5) 
among the 12 candidates reported in \citet{hogan08}. Hence,
the success rate of the technique employed by \citet{hogan08} to identify 
L dwarfs in the Hyades is larger than 58\%. \citet{hogan08} predicted two 
contaminants (photometric and proper motion nom-members) out of the 12 candidates, 
estimate confirmed by the presence of two early-M dwarfs in the sample.

%
%
\begin{figure*}
  \centering
  \includegraphics[width=0.49\linewidth, angle=0]{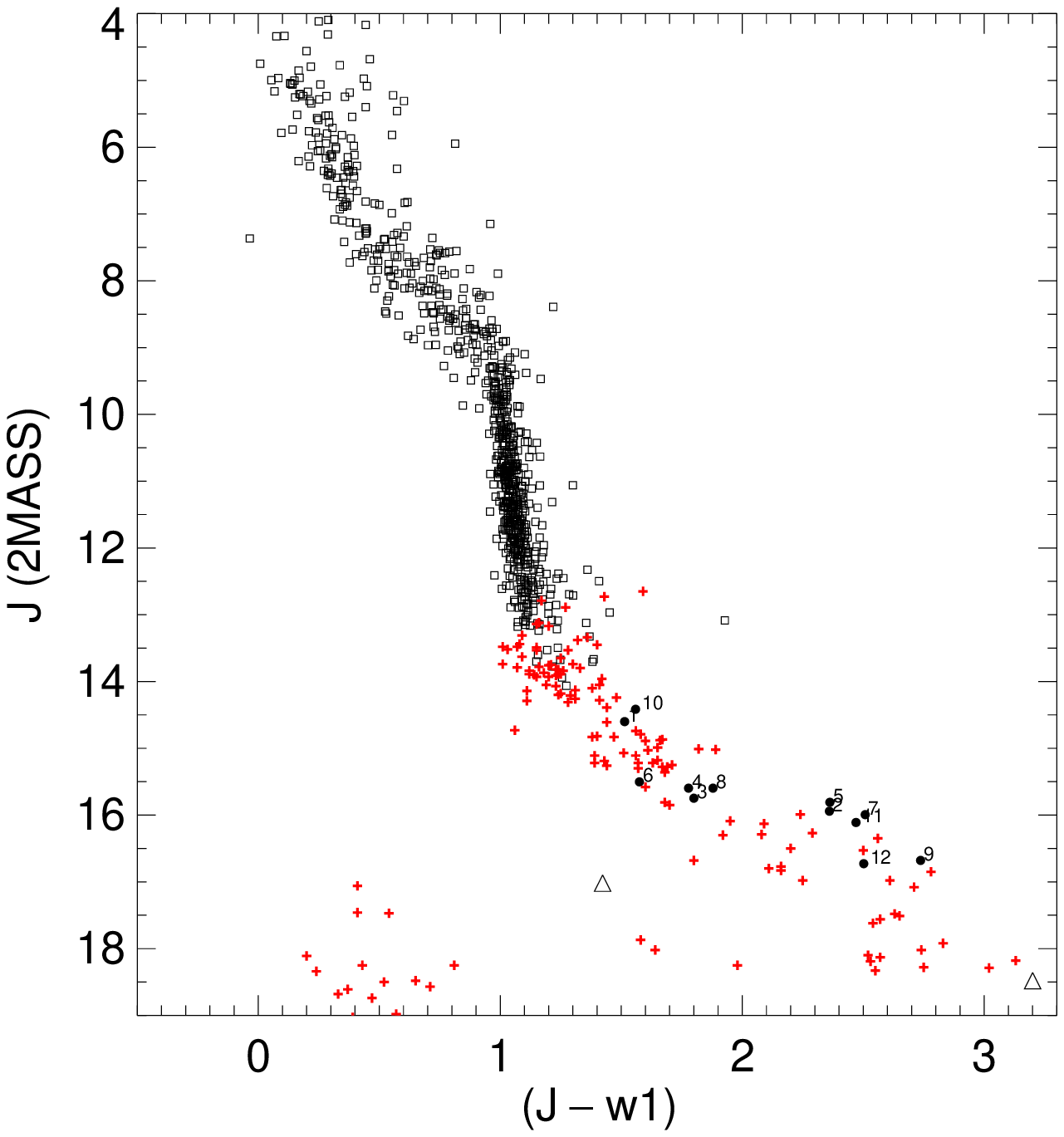}
  \includegraphics[width=0.49\linewidth, angle=0]{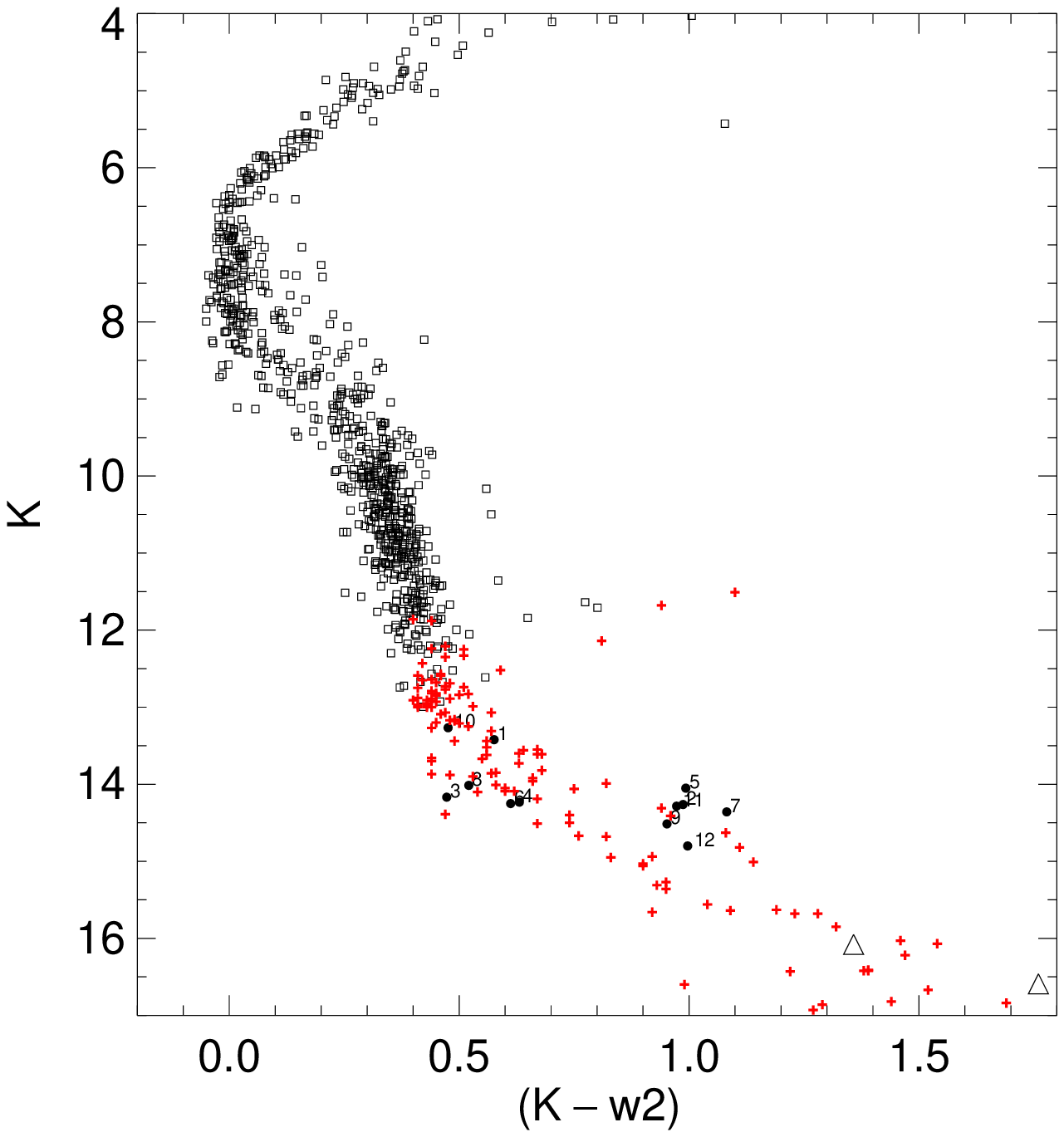}
  \caption{($J-w1$,$K$) and ($K_{s}-w2$,$K_{s}$) colour-magnitude diagrams for known 
Hyades stellar members \citep[open squares;][]{goldman13} and candidates from
\citet{hogan08} with optical spectra (dots marked with numbers following
the order in Table \ref{tab_Hyades_dL:log_obs}). Candidates Hya05 and Hya07
are rejected as members whereas the WISE photometry of candidate Hya02 is
affected by a close star, making its photometry and colours unreliable. 
Hyades T dwarfs from \citet{bouvier08a} are marked as open triangles.
Overplotted as red crosses are ultracool dwarfs with parallaxes from
\citet{dupuy12}, after removal of metal-poor dwarfs.
}
  \label{fig_Hyades_dL:CMD_WISE}
\end{figure*}
%

%
%
\begin{table*}
 \centering
 \caption[]{Near- and mid-infrared photometry, proper motions (in mas/yr),
spectral types with half a subclass uncertainty, spectroscopic distances 
(in pc) based on the $J$-band photometry, effective temperatures (in Kelvins)
using field L dwarfs with parallax measurements \citep{dahn02,vrba04}, 
masses (in M$_{\odot}$) from the BT-Settl models \citep{allard12},
and final membership assignment (Memb?)
for the 12 Hyades L dwarf candidates identified by \citet{hogan08}. The bottom
panels list the same datasets but for the two T dwarfs in \citet{bouvier08a}.
Note: Hya02 is a blend in WISE images. The WISE photometry of Hya04 and Hya10
is most likely affected by the contribution of a faint sources seen on the
UKIDSS GCS $K$-band images within the spatial resolution of WISE\@. \newline
}
 \begin{tabular}{c @{\hspace{2mm}}c @{\hspace{2mm}}c @{\hspace{2mm}}c @{\hspace{2mm}}c @{\hspace{2mm}}c @{\hspace{2mm}}c @{\hspace{2mm}}c @{\hspace{2mm}}c @{\hspace{2mm}}c @{\hspace{2mm}}c}
 \hline
 \hline
ID  & $J_{\rm LIRIS}$   &  $w1$  &  $w2$   & $\mu_{\alpha}\cos\delta$  & $\mu_{\delta}$  &  d & SpT & T$_{\rm eff}$ & Mass & Memb?  \cr
 \hline
    &  mag  &  mag   &  mag    & mas/yr     & mas/yr  & pc &  & K  & M$_{\odot}$ \cr
\hline
Hya01  &  ---   ( --- ) & 13.087 (0.024) & 12.846 (0.030) &   ---       &  ---         & 48.9$^{+3.8}_{-3.4}$  & M8.5 & 2379 (82) & 0.071--0.080 & Yes \\
Hya02  &  ---   ( --- ) & 13.582 (0.026) & 13.276 (0.032) &  123$\pm$10 & $-$28$\pm$8  & 57.7$^{+4.9}_{-5.0}$  & L1.5 & 2332 (114) & 0.068--0.079 & Y? \\
Hya03  & 15.647 (0.076) & 13.948 (0.029) & 13.695 (0.036) &  103$\pm$10 & $-$5$\pm$9   & 62.1$^{+4.7}_{-4.9}$  & L0.5 & 2320 (133) & 0.069--0.079 & Yes \\
Hya04  & 15.595 (0.044) & 13.819 (0.026) & 13.601 (0.035) &   70$\pm$10 & $-$25$\pm$11 & 67.1$^{+4.7}_{-4.9}$  & M9.5 & 2421 (185) & 0.073--0.097 & Yes \\
Hya05  & 15.729 (0.071) & 13.447 (0.025) & 13.058 (0.028) &  130$\pm$15 & $+$1.0$\pm$11 & 173.1$^{+4.0}_{-53.0}$ & M5.0 &  --- (---) &  ---  & No  \\
Hya06  & 15.462 (0.081) & 13.928 (0.028) & 13.639 (0.038) &   78$\pm$12 & $-$10$\pm$10 & 64.3$^{+4.5}_{-4.7}$  & M9.5 & 2421 (185) & 0.073--0.097 & Yes \\
Hya07  & 15.956 (0.073) & 13.486 (0.026) & 13.277 (0.032) &   94$\pm$15 & $-$40$\pm$15 & 443.8$^{+51.3}_{-115.7}$ & M3.0 &  --- (---) &  ---  & No  \\
Hya08  & 15.463 (0.044) & 13.718 (0.026) & 13.494 (0.033) &   89$\pm$9  & $-$15$\pm$9  & 57.9$^{+4.4}_{-4.6}$  & L0.5 & 2320 (133) & 0.069--0.079 & Yes \\
Hya09  &  ---   ( --- ) & 13.942 (0.028) & 13.563 (0.034) &    ---      &   ---        & 73.2$^{+5.5}_{-7.9}$  & L2.0 & 2368 (198) & 0.066--0.085 & Y? \\
Hya10  & 16.506 (0.059) & 12.857 (0.024) & 12.791 (0.027) &  115$\pm$11 & $-$12$\pm$11 & 82.6$^{+6.8}_{-6.6}$  & L1.0 & 2295 (82) & 0.068--0.075 & Y? \\
Hya11  &  ---   ( --- ) & 13.641 (0.025) & 13.311 (0.032) &  158$\pm$18 & $-$44$\pm$12 & 62.4$^{+6.0}_{-5.4}$  & L1.5 & 2332 (114) & 0.068--0.079 & Y? \\
Hya12  & 16.778 (0.069) & 14.226 (0.028) & 13.804 (0.047) &  101$\pm$11 & $-$15$\pm$11 & 57.3$^{+5.3}_{-5.8}$  & L3.5 & 1982 (114) & 0.054--0.063 & Y? \\
\hline
Hya20  &  ---   ( --- ) & 15.597 (0.048) & 14.722 (0.078) &    ---  &     ---    & 28.8$^{+0.3}_{-0.3}$  & T2.0 & 1361 (150) & 0.030--0.040 & Y? \\
Hya21  &  ---   ( --- ) & 15.280 (0.041) & 14.830 (0.100) &    ---  &     ---    & 55.5$^{+0.4}_{-0.1}$  & T1.0 & 1288 (150) & 0.030--0.040 & Yes \\
\hline
 \label{tab_Hyades_dL:table_results}
 \end{tabular}
\end{table*}
\subsection{WISE mid-infrared photometry}
\label{Hyades_dL:memb_phot_WISE}

We have also downloaded the Wide Field Infrared Survey Explorer (WISE) all-sky photometry 
\citep{wright10} to further assess the membership of these 12 L dwarf candidates. 
WISE observed the full sky in four mid-infrared filters centered at 3.4, 4.6, 12, and 22
microns down to 5$\sigma$ limits of 16.5, 15.5, 11.3, and 7.9 mag (Vega system) with
spatial resolutions of 6.1, 6.4, 6.5, and 12 arcsec, respectively.
We checked the WISE images where all candidates are detected in $w1$ and $w2$ with
signal-to-noise ratios in the 37--45 and 23--40, respectively. However, none of the 
L dwarfs are detected in the $w3$ and $w4$ bands (Table \ref{tab_Hyades_dL:table_results}).
We also downloaded the WISE photometry for the two T dwarfs, where the signal-to-noise
ratios in $w1$ and $w2$ are 22--26 and 10--14, respectively. As for the L dwarfs,
the T dwarfs are undetected in the other WISE passbands. The $K-w2$ and $J-w1$
colours of CFHT-Hy-20 (T2) and CFHT-Hy-21 (T1) are consistent with the typical
colours of field late-L and early-T dwarfs as depiacted in Fig.\ \ref{fig_Hyades_dL:CMD_WISE}.
We should mention that there is a faint nearby source on the GCS $K$-band image, 
which likely affects the WISE photometry of CFHT-Hy-21\@.

In Fig.\ \ref{fig_Hyades_dL:CMD_WISE} we display the ($J-w1$,$J$) and ($K_{s}-w2$,$K_{s}$) 
colour-magnitude diagrams for known members of the Hyades \citep[open squares;][]{goldman13}
to which we added the Hyades L and T dwarfs and field ultracool dwarfs with parallaxes
\citep*{dupuy12}. We observe two sub-groups among the 12 L dwarf
candidates. We find six candidates that lie below the Hyades sequence of low-mass stars 
and seem to bridge the gap with the Hyades T dwarfs. This group contains three late-M
dwarfs (Hya01, Hya04, and Hya06) and three L dwarfs (Hya03, Hya08, and Hya10).
These objects also fit the sequence of spectral types for ultracool dwarfs with 
parallaxes (red crosses in Fig.\ \ref{fig_Hyades_dL:CMD_WISE}), which is not too 
surprising because the age of the Hyades is close to the mean age of field dwarfs.

\begin{figure}
  \centering
  \includegraphics[width=\linewidth, angle=0]{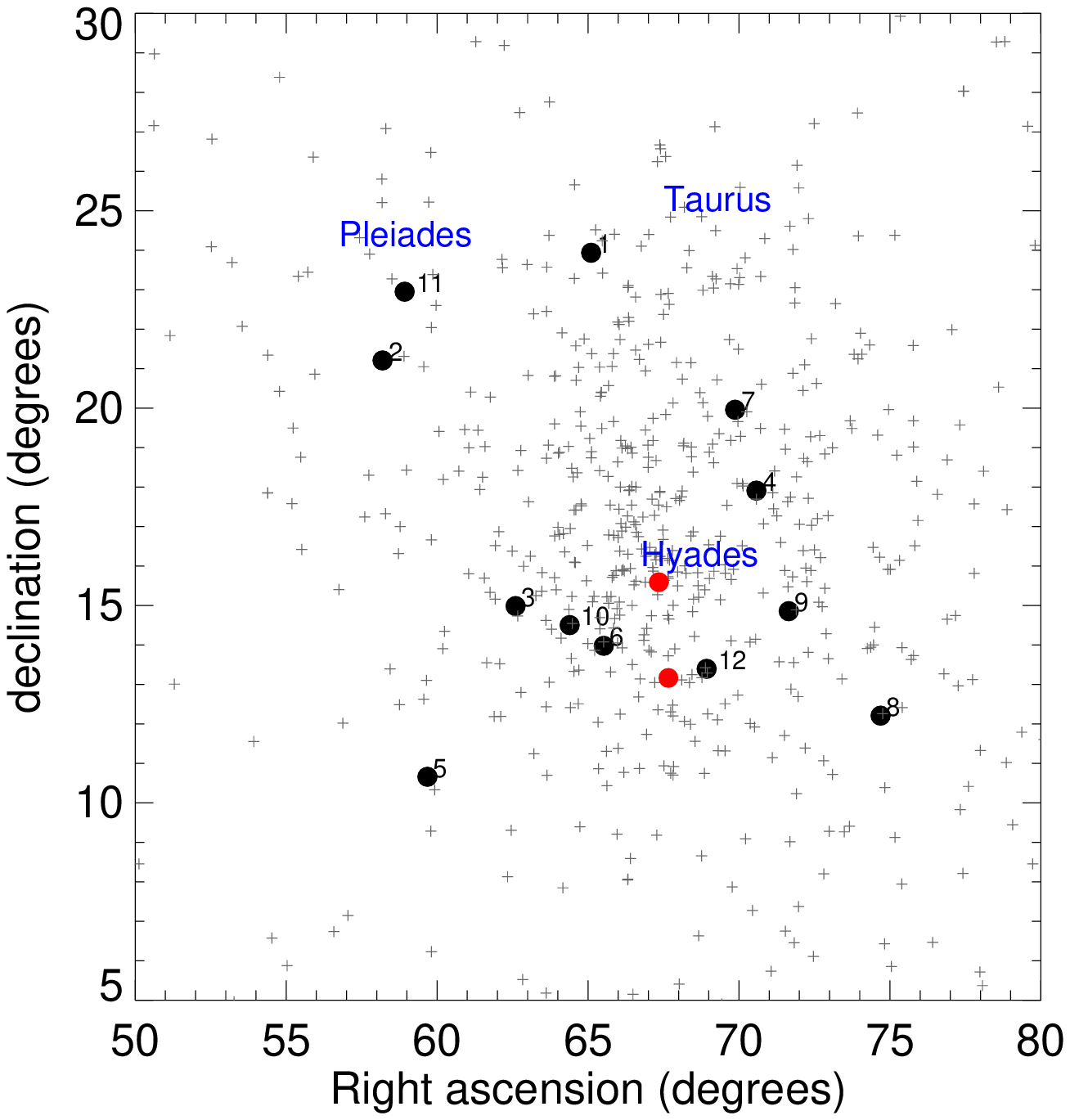}
  \caption{Spatial distribution of the L dwarf candidates from \citet{hogan08}
with their ID numbers from Table \ref{tab_Hyades_dL:log_obs}
and known members of the Hyades \citep[grey crosses;][]{goldman13}.
Overplotted as red dots are the two Hyades T dwarfs from \citet{bouvier08a}.
The approximate centres of the Pleiades, Taurus, and Hyades regions are
also labelled for reference.
}
  \label{fig_Hyades_dL:plot_ra_dec}
\end{figure}

The group showing the reddest colours in $J-w1$ and $K_{s}-w2$ 
(Fig.\ \ref{fig_Hyades_dL:CMD_WISE}) includes two
spectroscopic non-members (Hya05 and Hya07; see Section \ref{Hyades_dL:memb_spectro})
and Hya02 which has a companion on the UKIDSS GCS $K$-band image unresolved on the
WISE images due to its poorer spatial resolution at 3.6 and 4.5 microns 
\citep[$\sim$6.5 arcsec;][]{wright10}.
For this reason, we urge caution in the interpretation its magnitudes and colours.
The other three sources in this `red' group are Hya09, Hya11, and Hya12, which we
classify as L dwarfs (Table \ref{tab_Hyades_dL:table_results}; Section \ref{Hyades_dL:memb_spectro}).
One of them, Hya11, has a position in the sky very close to the Pleiades 
(Fig.\ \ref{fig_Hyades_dL:plot_ra_dec})
and is reported as a non-member in Table A1 of \citet{lodieu12a} because 
of its large proper motion 
($\mu_{\alpha}cos\delta$\,=\,161.1$\pm$2.3 and $\mu_{\delta}$\,=\,$-$44.5 $\pm$2.3 mas/yr),
which is consistent with the Hyades proper motion. \citet{casewell08a} and
\citet{faherty09} reported proper motions of (172,$-$21) mas/yr and (142,$-$25) mas/yr
for Hya11, respectively. 
We also note that no companion was detected around
Hya11 down to magnitude limits of $J$\,=\,20.5 mag and $K$\,=\,18.5 mag with
separations in the 2--31 arcsec range \citep{allen07}.
The other two sources remain with mid-infrared excess and could be younger than the
Hyades but we do not detect H$\alpha$ in emission or see obvious signs of youth 
(e.g.\ weaker atomic lines) at the resolution of our spectra. Nonetheless, we
cannot discard at this stage that these mid-IR excesses characteristics of youth 
may extent to the typical age of the Hyades and hence, could be used as a criterion
to confirm their membership.

%
%
%

%
%
%
\section{Physical properties of Hyades L dwarf candidates}
\label{Hyades_dL:physical_properties}
\subsection{Spectroscopic distances}
\label{Hyades_dL:properties_dist}

To further assess the likelihood of the new late-M and L dwarfs to belong to the
Hyades cluster, we have computed spectroscopic distances based on the $J$-band
absolute magnitude vs spectral type relationship from \citet*{dupuy12}, which
is valid for field (old) ultracool dwarfs with spectral types later than M6 
with a rms of 0.4 mag.

For the three late-M dwarfs (Hya01, Hya04, and Hya06), we derive distances of
48.9$^{+3.8}_{-3.4}$ pc, 67.1$^{+4.7}_{-4.9}$ pc, and 64.3$^{+4.5}_{-4.7}$ pc, 
respectively (Table \ref{tab_Hyades_dL:table_results}).  
Assuming a mean distance of 46.34$\pm$0.27 pc for the Hyades \citep{perryman98}, 
Hya01 is bona-fide member whereas the other two candidates lie slightly further 
away. However, we cannot discard these objects as members at that stage for several
reasons. First, the halo of the Hyades is estimated to be twice the tidal 
radius \citep[$\sim$10 pc;][]{perryman98,goldman13}. Second, the 
typical uncertainty on our spectroscopic distances is of the order of 10\% 
considering our uncertainty of half a subclass on the spectral type to which
we should add another 20\% error due to the uncertainty of 0.4 mag in the
spectral types vs absolute magnitude relation of \citet*{dupuy12}.
Third, we cannot discard the effect of mass segregation due to dynamical 
evolution for ages older than 600 Myr.

For the L dwarfs, we derive spectroscopic distances in the range 57--67 pc, 
suggesting that most candidates lie roughly at the distance of the Hyades. 
The only two exceptions are Hya09 and Hya10 which lie further than 70 pc,
73$^{+5.5}_{-7.9}$ pc and 82.6$^{+6.8}_{-6.6}$ pc, respectively (Table \ref{tab_Hyades_dL:table_results}) 
but, again, we cannot discard these objects as members at this stage as pointed 
out above. We classify L dwarfs within the halo of the Hyades as members (``Yes'')
while the other remain possible members (``Y?''; Table \ref{tab_Hyades_dL:table_results}).
This effect might be a result of the dynamical evolution suffered by the cluster
\citep[e.g.][]{vesperini97,lamers10}.
However, we caution that our interpretation is based on absolute magnitude-spectral type
relationships valid for field dwarfs, which might not be adequate because the Hyades
cluster is on average younger than typical nearby ultracool dwarfs.

We note that the two candidates from \citet{hogan08} classified as M3 (Hya07) and M5 (Hya05)
dwarfs would lie at distances of $\sim$169--226 and $\sim$390--560 pc. They lie 
clearly beyond the Hyades cluster and are background contaminants.

We applied the same absolute magnitude vs spectral type relationship to the
two T dwarf members proposed by \citet{bouvier08a}. We inferred spectroscopic
distances of 28.8$\pm$0.3 pc and 55.5$\pm$0.4 pc for CFHT-Hy-20 and CFHT-Hy-21, respectively,
based on their $J$-band photometry. These estimates would place those T dwarfs
within the halo of the Hyades assuming a mean distance of 46 pc \citep{perryman98}.
The difference in $J$-band absolute magnitude between a field L3 dwarf and 
a T1 dwarf is about two magnitudes \citep[e.g.][]{dupuy12}, suggesting
these T dwarfs as bona-fide members of the Hyades cluster.
We suggest to initiate a parallax program on the Hyades L and T dwarfs
to place a more reliable constraint on their distance.

\begin{figure}
  \centering
  \includegraphics[width=\linewidth, angle=0]{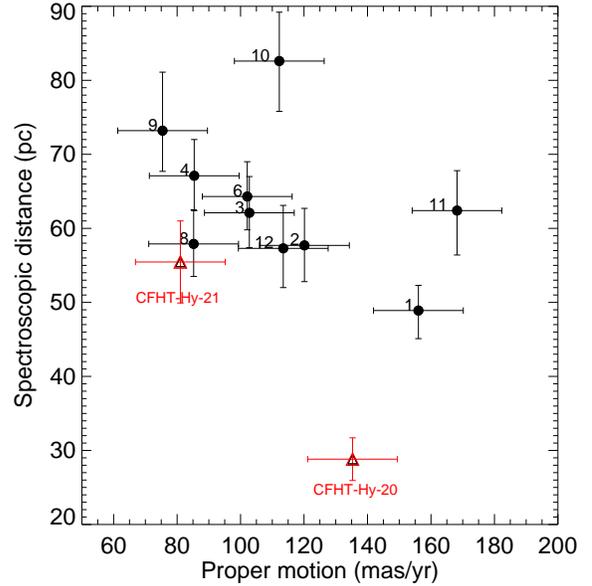}
  \caption{Spectroscopic distances (pc) vs total proper motion (mas/yr)
for the Hyades L (black filled dots with their ID numbers from 
Table \ref{tab_Hyades_dL:log_obs}) and T (red open triangles) dwarfs.
Error bars of 14 mas/yr are assumed for the total proper motion
whereas uncertainties on the spectroscopic distances are from
Table \ref{tab_Hyades_dL:log_obs}.
}
  \label{fig_Hyades_dL:spec_dist_vs_PM}
\end{figure}

We should point out that distance might has an important bearing on the proper 
motion because an object at the back of the cluster (d\,$\sim$60 pc) as
opposed to the front (d\,$\sim$30 pc) would have a noticeably smaller
motion. In Fig.\ \ref{fig_Hyades_dL:spec_dist_vs_PM} we display
the spectroscopic distances as a function of the total proper motion
for the spectroscopic L dwarfs and the two Hyades T dwarfs.
The typical error bars of the proper motion in each axis derived from the cross-match
between the UKIDSS GCS and 2MASS are of the order of 10 mas/yr \citep{lodieu07a},
in agreement with the values of $\pm$7 mas/yr reported by \citet{hogan08}.
The same uncertainty is used for the T dwarfs \citep{bouvier08a}.
The typical uncertainties on the spectroscopic distances are $\sim$10\%.
We observe a linear relation defined by seven/eight of the ten L dwarfs,
(Hya01--09 and Hya12, possibly Hya11 as well) which might argue against 
cluster membership for Hya10 and one of the two T dwarfs (CFHT-Hy-20; T2).

%
%
\subsection{Effective temperatures}
\label{Hyades_dL:properties_Teff}

In the following two sub-sections, we assume that all candidates are members
of the Hyades cluster, except Hya05 and Hya07\@.
We estimated the effective temperatures of the spectroscopic M and L dwarfs
in the Hyades, using the mean temperature derived from the spectral type 
vs effective temperature relations of field L dwarfs presented in 
\citet{dahn02} and \citet{vrba04}. The coolest L dwarf is Hya12 with an
effective temperature below 2000\,K\@. We assumed
that these relationships are valid for old field L dwarfs are applicable to the
Hyades. This assumption is not exactly correct because members of the Hyades
have an age of 575--675 Myr while field L dwarfs are expected to be older than
1 Gyr. \citet{luhman99a} and \citet{luhman03b} proposed a temperature scale
for young brown dwarfs in IC\,348, where the difference between 3 Myr-old
M dwarfs and their field counterpart is less than 200--400\,K\@.
Hence, we do not expect a difference larger than 100\,K between old field
dwarfs and Hyades members because of the smaller age difference. This
estimate is in line with model predictions \citep{baraffe98,allard12}.
Our adopted effective temperatures are listed in Table \ref{tab_Hyades_dL:table_results}
and likely represent upper limits. The uncertainties on the temperatures
represent the sum in quadrature of the dispersion observed for a given spectral 
type listed in the tables of \citet{dahn02} and \citet{vrba04} and our uncertainty
of half a sub-type on the spectral type.

\subsection{Masses}
\label{Hyades_dL:properties_Mass}

We derived masses for the confirmed Hyades M and L dwarfs using the latest
BT-Settl models \citep{allard12}. We assumed an age of 625 Myr for the Hyades.
We converted the effective temperatures derived from the optical spectral
types into masses using the BT-Settl models, taking into account the
uncertainties on the temperatures.
We list the range in masses in Table \ref{tab_Hyades_dL:table_results}.

\subsection{Lithium}
\label{Hyades_dL:properties_Li}

According to theoretical evolutionary models, L dwarfs are either very low mass 
stars with ages of a few Gyr or brown dwarfs with masses close to the
hydrogen-burning mass limit or younger than 1 Gyr \citep{burrows97,chabrier00a}.
The Li element is rapidly destroyed in the interior of stars, on time scales 
shorter than $\sim$\,150\,Myr, and in massive brown dwarfs on time scales of a few Gyr.
Brown dwarfs with masses lower than 0.055--0.060 M$_{\odot}$ do not burn this 
element in their interiors because their central temperature is not hot enough 
to produce this fusion reaction \citep{dantona94,ushomirsky98,chabrier00a}
According to our mass estimates, the majority of our candidates in the Hyades 
cluster are very low mass stars and hence, we expect that these objects have 
destroyed their Li content. However our candidate with the latest spectral
type (Hya12; L3.5) has an estimated mass of only 0.054-0.063 M$_{\odot}$, hence 
it is expected that it has preserved most of its Li. Given the poor signal-to-noise 
and resolution of our spectrum, we can only impose an upper limit of 1\,\AA{} 
on the absorption induced by the lithium line at 6708\,\AA{}. 

In order to confirm this claim, we should obtain a higher resolution spectrum 
with higher signal-to-noise to detect lithium in absorption at 6708\,\AA{} 
and apply the lithium test \citep{magazzu91,magazzu92,rebolo92,martin94b}.
We place the Hyades T dwarfs in the substellar regime with masses in the
0.03--0.04 M$_{\odot}$ range, estimates slightly lower than the 0.05 M$_{\odot}$ 
quoted by \citet{bouvier08a} using the DUSTY models \citep{chabrier00c}.

%
%
\section{The Hyades mass function}
\label{Hyades_dL:MF}

Before discussing the shape of the Hyades (system) mass function, we need to compile
a sample of members as complete and unbiased as possible. \citet{goldman13}
published a sample of 776 members with masses ranging between 2.6 and 
0.11 M$_{\odot}$ out to 30 pc from the cluster centre. Our spectroscopic
follow-up confirms three late-M (Hya01, Hya04, and Hya06) and 
two early-L dwarfs (Hya03 and Hya08) with masses between
0.1 and 0.069 M$_{\odot}$ within an circular area centered on (RA,dec)\,=\,(67,12) 
degrees and a radius of 11 degrees \citep[Figure 1 in][]{hogan08}.
We do not include the other spectroscopic L dwarfs in this discussion because
we are unsure about the membership to the Hyades at this stage.
We counted 379 sources with masses less than 1 M$_{\odot}$ in the survey 
of \citet{goldman13} within the 11-degree radius covered by \citet{hogan08}. 
We updated the Hyades mass function by adding the five spectroscopically-confirmed 
Hyades members to the sample of \citet{goldman13}. We estimate our results to
be valid for magnitudes brighter than $K$\,$\sim$\,15 mag, corresponding to
masses of 0.05 M$_{\odot}$ according to the 600 Myr isochrones of the BT-Settl models.

At lower masses, we argue for the membership of one of the T dwarfs (CFHT-Hy-21) 
reported by \citet{bouvier08a} whereas the other T dwarf (CFHT-Hy-20) remains as a 
candidate, as is the latest L dwarf in our sample (Hya12). We do not add a bin in 
the substellar regime to our revised mass function due to the large uncertainties
associated with the membership of the coolest candidates. We should mention 
that the two T dwarfs from \citet{bouvier08a} were discovered in a $\sim$16 
square degree area with inhomogeneous coverage accross the Hyades cluster. 
These authors argued that there could be 10--15 brown dwarfs down to 
0.013 M$_{\odot}$ although the original population of the cluster may have 
been of the order of 100--150 prior to the impact of dynamical evolution.

%
%
\begin{figure}
  \centering
  \includegraphics[width=\linewidth, angle=0]{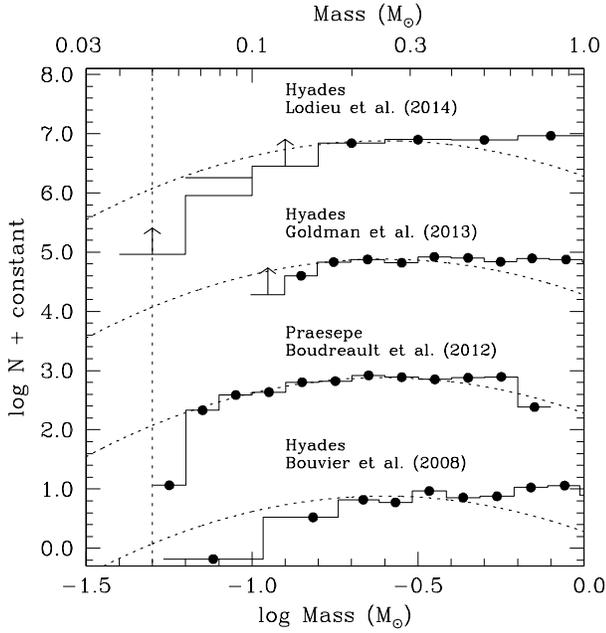}
  \caption{Hyades mass function (number of objects per interval of mass in
logarithmic units; lower limits represented as arrows or two horizontal bars 
in the case of the second lowest mass bin) derived from this work, 
combining the results of
\citet{goldman13} with our spectroscopic follow-up. For comparison, we show the
mass function of Praesepe (bottom) derived from the photometric and astrometric 
analysis of the UKIDSS GCS \citep{boudreault12}, the original mass functions
of \citet{goldman13} and \citet{bouvier08a}. Overplotted as a dashed line is the
the field mass function \citep{chabrier05a}. We normalised all the MFs to the 
log-normal fit of \citet{chabrier05a} at $\sim$0.3\,$M_{\odot}$ ($\log (M)$\,$\sim$\,$-$0.5).
The vertical dashed line represent the completeness limit of our work.
}
  \label{fig_Hyades_dL:MF_Hyades}
\end{figure}
%

%
%
\begin{table}
 \centering
 \caption[]{Number of objects (dN) per mass bin in logarithmic units (d$\log(M)$)
for the Hyades (this work) and Praesepe \citep{boudreault12}.
Each mass bin has a width of 0.1 in logarithmic space.
The last two bins represent the numbers of late-M, L, and T dwarfs most 
likely members of the Hyades confirmed in this work and \citet{bouvier08a}
whereas the higher mass bins represent members from \citet{goldman13} 
located in the area covered by \citet{hogan08}.
}
 \begin{tabular}{@{\hspace{0mm}}l @{\hspace{1mm}}c @{\hspace{1mm}}c @{\hspace{1mm}}c @{\hspace{1mm}}c @{\hspace{1mm}}c @{\hspace{1mm}}c @{\hspace{1mm}}c@{\hspace{0mm}}}
 \hline
d$\log(M_{\rm mean})$ &   $-$0.10  & $-$0.30  & $-$0.50   & $-$0.70  & $-$0.90 & $-$1.10 & $-$1.30 \cr
 \hline
dN$_{\rm Hyades}$  & 101 & 86$\pm$9 & 87$\pm$9  & 76$\pm$9 & $>$31   & 5--10    &  $>$1   \cr
 \hline
dN$_{\rm Prae}$    &   ---      &  267$\pm$11 & 257$\pm$16  & 255$\pm$6 & 185$\pm$16  &  104$\pm$17 &   $>$2     \cr
dN$_{\rm Prae}$ scaled & ---    & 90$\pm$4 & 87$\pm$5  & 86$\pm$2 & 63$\pm$5  &  35$\pm$6 &   $>$0.7     \cr
\hline
 \label{tab_Hyades_dL:table_MF}
 \end{tabular}
\end{table}

\citet{goldman13} presented the Hyades mass function down to 0.11 M$_{\odot}$
although the last two bins in their figure 14 are incomplete ($\log(M)$\,=\,$-$0.85
and $-$0.95 with a width of 0.1), corresponding to the sixth column in
Table \ref{tab_Hyades_dL:table_MF}. In Table \ref{tab_Hyades_dL:table_MF}, we
quote the numbers of members per bin of 0.2 M$_{\odot}$ in logarithmic
units between 1 and 0.1 M$_{\odot}$ from \citet{goldman13} located in the
area covered by \citet{hogan08}. In the $\log(M)$\,=\,$-$1.1 bin, we added 
the three late-M and two early-L dwarfs confirmed as the most probable Hyades
members. This is likely a lower limit (arrow in Fig.\ \ref{fig_Hyades_dL:MF_Hyades})
because some of the spectroscopic L
dwarfs reported in this work might turn out to be members once we have
collected additional data. If we include all 10 late-M and early-L
dwarfs, the upper limit of the second lowest-mass bin would still lie below 
the field mass function. We have only considered CFHT-Hy-21 in the last 
bin ($\log(M)$\,=\,$-$1.3; Table \ref{tab_Hyades_dL:table_MF}) but it 
should be seen as a lower limit mainly because the area covered by 
\citet{bouvier08a} is much smaller than the other two surveys considered 
in our analysis (arrow in Fig.\ \ref{fig_Hyades_dL:MF_Hyades}). 
To conclude, we should highlight that the shape of the 
Hyades substellar mass function still remains to be determined with much 
better accuracy. Moreover, we emphasise that we do not correct the mass bins 
for binaries and point out that our analysis makes use of different catalogues
coming from surveys with distinct depths and completeness.

Bearing in mind the aforementioned caveats, we compare the Hyades mass function 
to Praesepe (Table \ref{tab_Hyades_dL:table_MF}; Fig.\ \ref{fig_Hyades_dL:MF_Hyades}),
as published by \citet{boudreault12} from an photometric and astrometric
survey using the UKIDSS GCS comparable to the study of \citet{goldman13}.
The last bin of the Praesepe mass function is highly incomplete and represent
a lower limit although we note that we have now confirmed spectroscopically 
a L dwarf member with GTC/OSIRIS \citep[, UGCS\,J084510.66$+$214817.1 (L0.3$\pm$0.4);][]{boudreault13}.
The last row in Table \ref{tab_Hyades_dL:table_MF} quotes the number of
objects for Praesepe scaled to the Hyades mass function (dN$_{\rm Prae}$ scaled) 
around $\log(M)$\,=\,$-$0.5\@, which corresponds to the peak in the field mass 
function \citep{chabrier03,chabrier05a}.
We find that the number of Hyades members (5--10) in the mass bin centered on
0.08 M$_{\odot}$ is lower than the values of Praesepe ($\sim$3 sigma or more), 
a possible result of dynamical evolution.
We note that the last bin cannot be directly compared because the
UKIDSS GCS survey of \citet{boudreault12} is not sensitive to late-L
and T dwarfs in Praesepe and the survey by \citep{bouvier08a} probes
a very small area of the Hyades.

%
%
\section{Conclusions}
\label{Hyades_dL:conclusions}

We have presented a photometric, astrometric, and spectroscopic follow-up of 12 L dwarf
candidates in the Hyades cluster announced by \citet{hogan08}. We have also presented
mid-infrared photometry from WISE for all L dwarf candidates and the two 
previously-published T dwarfs. We can summarise the main results of our study as follows:
\begin{itemize}
\item[$\bullet$] we classify optically three L dwarf candidates as late-M dwarfs 
(Hya01, Hya04, and Hya06) with spectral types between M8.5 and M9.5,
which are bona-fide members of the Hyades cluster
\item[$\bullet$] we confirm the L dwarf status of seven candidates, with spectral types in the 
L0--L3.5 range, two of them being very likely members of the Hyades Hya03 and Hya08\@. 
Our GTC OSIRIS optical spectra of these Hyades L dwarf members represent spectral templates at an 
age of 625 Myr, which will we make publicly available to the community
\item[$\bullet$] we classify the remaining two candidates as M3 (Hya07) and M5 (Hya05),
rejecting them as Hyades members
\item[$\bullet$] we find infrared excesses in four L dwarfs (Hya02, Hya09, Hya11, and Hya12)
confirmed spectroscopically, which are typical of young L dwarfs. This result suggests that 
these objects might belong to younger star-forming regions like Taurus or the Pleiades cluster 
or the these features extent to the age of the Hyades cluster
\item[$\bullet$] we estimate spectroscopic distances of 50--90 pc for the seven
candidates confirmed spectroscopically as L dwarfs (Hya02, Hya03, Hya08--12), using the 
latest absolute $J$-band magnitude vs spectral type relationship of nearby ultracool dwarfs
\item[$\bullet$] we derive effective temperatures in the range 1980--2420\,K, based
on the spectral type vs temperature relation of nearby old field dwarfs
\item[$\bullet$] we infer masses in the 0.08--0.054 M$_{\odot}$ from the latest BT-Settl models,
assuming an age of 625 Myr for the Hyades L dwarfs, placing the coolest object below
the stellar/substellar boundary if a true member of the cluster
\item our mass function is in line with a possible deficit of very low-mass stars and brown dwarfs 
in the Hyades, an effect likely due to dynamical evolution
\end{itemize}

The original selection of \citet{hogan08} was made in a large area covered by
an early data release of the UKIDSS GCS but limited by the 2MASS depth,
reaching $K$\,$\sim$\,15 mag. We aim at extending this
search to look for fainter and cooler L and T dwarfs in two ways. On the one hand,
we plan to cross-match WISE with the UKIDSS GCS $K$-band survey to exploit the full
depth of 15.5. mag of the WISE $w2$ band and the 100\% completeness of the GCS down
to $K$\,$\sim$\,18 mag (Lodieu et al.\ in prep). On the other hand, we will complement that
search by cross-correlating the UKIDSS GCS $K$-band with the upcoming UKIRT Hemisphere 
Survey $J$-band observations whose depth is similar to the UKIDSS Large Area Survey
\citep{dye06,warren07a}, as we did for the Pleiades \citep{lodieu12a},
$\alpha$\,Persei \citep{lodieu12c}, and Praesepe \citep{boudreault12} clusters.
Moreover, a moderate-resolution spectroscopic survey should be carried out to
investigate the radial velocity of the L dwarf confirmed spectroscopically whose
distances lie at the lower and upper limits of the tidal radius of the Hyades cluster.
These observations will complement the high precision that the Gaia mission will
provide down to a spectral type of $\sim$M8 in the Hyades \citep{deBruijne14}.

%
%
\section*{Acknowledgments}
NL was funded by the Ram\'on y Cajal fellowship number 08-303-01-02
and the national program AYA2010-19136 funded by the Spanish ministry of 
Economy and Competitiveness (MINECO). NL thanks Jos\'e Acosta Pulido for his help 
and support with the data reduction of the LIRIS datasets.
VJSB is funded by the MINECO project AYA2010-2053\@. We thank Bartosz Gauza
for his help with the Osiris observations. NL thanks Paul Dobbie and 
Nigel Hambly for valuable comments on the original version of the paper.

This work is based on observations made with the Gran Telescopio Canarias
(GTC), operated on the island of La Palma in the Spanish Observatorio del
Roque de los Muchachos of the Instituto de Astrof\'isica de Canarias under
CAT programmes GTC27-13B and GTC65-13B\@.

Based on observations made with the William Herschel Telescope operated on 
the island of La Palma by the Isaac Newton Group in the Spanish Observatorio 
del Roque de los Muchachos of the Instituto de Astrof\'isica de Canarias under 
the CAT programme 158-WHT38/13B\@.

This research has made use of the Simbad and Vizier \citep{ochsenbein00}
databases, operated at the Centre de Donn\'ees Astronomiques de Strasbourg 
(CDS), and of NASA's Astrophysics Data System Bibliographic Services (ADS). 

GAIA (Graphical Astronomy and Image Analysis Tool) is a derivative of the 
Skycat catalogue and image display tool, developed as part of the Very Large
Telescope project at European Southern Observatory. Skycat and GAIA are free 
software under the terms of the GNU copyright.

%
%
\bibliographystyle{mn2e}
\bibliography{../../AA/mnemonic,../../AA/biblio_old}

\bsp

\label{lastpage}

\end{document}